\documentclass[10 pt]{iopart}
\usepackage{graphicx}
\usepackage{amssymb}
\usepackage{float}
\usepackage{epsfig}
\usepackage{soul}
\usepackage{cite}
\usepackage{color,soul}
\usepackage{titlesec}
\graphicspath{ {./figs/} }
\binoppenalty=\maxdimen 
\relpenalty=\maxdimen

\newcommand{\figref}[2][]{(fig.~\ref{#2}#1)}

\begin{document}
\title[Soft interactions and bound mobility in diffusion in crowded
environments]{Effects of soft interactions and bound mobility on
  diffusion in crowded environments: a model of sticky and slippery
  obstacles} \author{ Michael W. Stefferson$^1$, Samantha
  A. Norris$^1$, Franck J. Vernerey$^2$, Meredith D.  Betterton$^1$,
  and Loren E. Hough$^1$ } \address{$^1$ Department of Physics,
  University of Colorado, Boulder \\ $^2$ Department of Mechanical
  Engineering, University of Colorado, Boulder }
\ead{loren.hough@colorado.edu}
\begin{abstract}
  Crowded environments modify the diffusion of macromolecules,
  generally slowing their movement and inducing transient anomalous
  subdiffusion. The presence of obstacles also modifies the kinetics
  and equilibrium behavior of tracers.  While previous theoretical
  studies of particle diffusion have typically assumed either
  impenetrable obstacles or binding interactions that immobilize the
  particle, in many cellular contexts bound particles remain mobile.
  Examples include membrane proteins or lipids
  with some entry and diffusion within lipid
  domains and proteins that can enter into membraneless organelles or 
  compartments such as the nucleolus.  Using a lattice model, we
  studied the diffusive movement of tracer particles which bind to
  soft obstacles, allowing tracers and obstacles to occupy the same
  lattice site. For sticky obstacles, bound tracer particles are
  immobile, while for slippery obstacles, bound tracers can hop
  without penalty to adjacent obstacles. In both models, binding
  significantly alters tracer motion. The type and degree of motion
  while bound is a key determinant of the tracer mobility: slippery
  obstacles can allow nearly unhindered diffusion, even at high
  obstacle filling fraction. To mimic compartmentalization in a cell,
  we examined how obstacle size and a range of bound diffusion
  coefficients affect tracer dynamics. The behavior of the model is
  similar in two and three spatial dimensions.  Our work has
  implications for protein movement and interactions within cells.
\end{abstract}
\maketitle
\ioptwocol 
	
\section{ Introduction }

The diffusion of macromolecules in crowded environments is generally
slowed relative to the uncrowded case, and particle motion can undergo
transient anomalous subdiffusion \cite{hofling_anomalous_13}.  The
motion of lipids or macromolecules within biological membranes can be
affected by
crowding \cite{schutz_singlemolecule_97,schutz_properties_00, nicolau_sources_07, weigel_ergodic_11, javanainen_anomalous_13, krapf_chapter_15, jeon_protein_16, sadegh_plasma_17},
because the membrane contains both macromolecules and inhomogeneities
in membrane composition \cite{petropoulos_membrane_90,
  veatch_organization_02}. In the cell interior, macromolecules,
organelles and other cellular structures can inhibit motion, 
or in contrast, enhance sampling of non-crowded regions \cite{stylianidou_cytoplasmic_14}.
Biological crowders can also contain interaction sites which further
modify the macromolecular motion \cite{crowley_protein_11}.  The
kinetics and equilibrium behavior of interactions between mobile
proteins can be modified by crowding \cite{mourao_connecting_14,
  zhdanov_kinetics_15}. The magnitude of the effects of crowding on
macromolecular motion and reactions is important to determine the
limiting rate of biological processes such as signaling receptor
activation.
  
Because of its biological importance, the effects of crowding on
diffusion and macromolecular interactions have seen significant
experimental and theoretical work \cite{dix_crowding_08,
  dirienzo_probing_14}. Lattice gas models have been used to
demonstrate the effects of crowding \cite{saxton_lateral_87,
  saxton_anomalous_94}, binding \cite{saxton_anomalous_96,
  berry_monte_02}, and repulsion \cite{ellery_analytical_16} on the
diffusion of tracer particles. These effects---including transient
anomalous diffusion at short times and hindered normal diffusion at
long times---have been studied for both immobile
\cite{ghosh_nonuniversal_15} and mobile obstacles
\cite{wedemeier_how_09, berry_anomalous_14}.

Although most theoretical work has focused on anomalous diffusion in
crowded systems made up of impenetrable obstacles with attractive or
repulsive surfaces \cite{saxton_lateral_87, saxton_anomalous_94,
  saxton_anomalous_96, ellery_characterizing_14,
  ellery_analytical_16}, there is growing evidence of the importance
of soft compartments and barriers in biological systems.  In
membranes, lipids can be only partially excluded from lipid rafts or
domains. When they do interact, they can still diffuse within them
\cite{fujiwara_phospholipids_02, forstner_attractive_08,
  ehrig_nearcritical_11, silvius_partitioning_05}.  Lipid motion can
be hindered, though not stopped, near $\alpha$-synuclein protein
aggregates \cite{iyer_membranebound_16}. For all of these cases,
theoretical considerations of a two-dimensional system should include
the effects of the soft interaction potentials and bound-state
mobility.

Inside the cell, intrinsically disordered or low-complexity domains
can act as soft obstacles or wells, with rapid diffusion within the
wells.  Membrane-less organelles spontaneously form from
low-complexity domain proteins. They are typically highly dynamic
assemblies \cite{brangwynne_germline_09}, which show fast
intra-particle diffusion times, and allow rapid entry and exit of
constituents \cite{molliex_phase_15}.  Proteins which interact with
intrinsically disordered proteins can still diffuse during the binding
interaction \cite{hough_molecular_15, raveh_slideandexchange_16}.
This effect may be particularly pronounced in the central channel of
the nuclear pore complex, which contains a high density of binding
sites on intrinsically disordered domains.  Recent simulation work
suggests that the disordered protein binding pockets can exchange on
transport factors \cite{raveh_slideandexchange_16}, providing a clear
mechanism for mobility while bound to an obstacle.  Particles are
weakly excluded from individual disordered protein chains due to the
lowering of the polymer chain entropy \cite{timney_simple_16}, but are
expected to allow other macromolecules to enter, and pass through,
the space that they occupy.  The increasingly recognized importance of
proteins which are intrinsically disordered or contain low-complexity
domains within their assemblies warrants a more careful consideration
of the differences between the previously well-studied models, in
which binding immobilizes the bound species, and a model which
includes soft interactions and obstacles or barriers in which the
bound species may remain mobile.

\begin{figure}[t]
	\centering
  \includegraphics[width=0.5\textwidth]{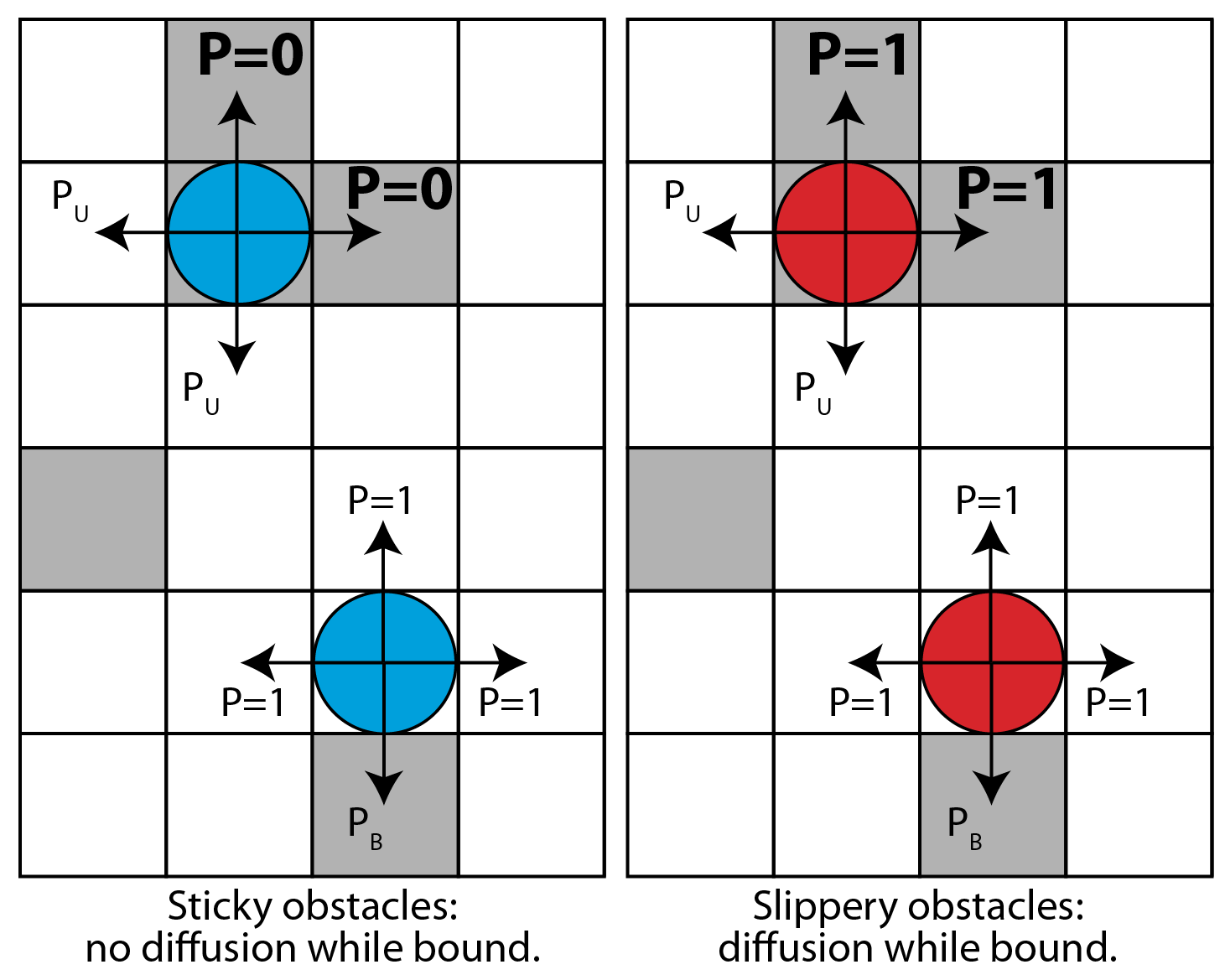}
  \caption{Model schematic. Tracers (colored circles) hop on a lattice
    of empty sites (white squares) and obstacles (gray
    squares). Tracer binding with a soft interaction potential allows
    them to overlap with obstacles (top). For sticky obstacles, the
    only allowed moves of a bound tracer are to empty sites
    (unbinding).  For slippery obstacles, tracers can hop to other
    obstacles while remaining bound.  Arrows denote possible moves and
    P the probability that a given move is accepted.}
  \label{fig:modelcartoon}
\end{figure}

Motivated by the biological importance of binding interactions which
can retain mobility of the bound particle, we studied a minimal model
with bound tracer mobility (figure \ref{fig:modelcartoon}). In our
model, tracer particles move on a 2D or 3D lattice in the presence of
immobile obstacles, to which the tracers can bind.  A primary
distinction between our model and many others that consider binding or
adhesion is that others typically consider adhesion between a tracer
and an adjacent hard obstacle, in which there is no overlap between
tracers and a hard obstacles core \cite{saxton_anomalous_96,
  wedemeier_how_09, ellery_analytical_16}.  Here, obstacles are soft:
tracer particles can overlap with obstacles, with an energy penalty
(or gain) $\Delta G$ upon moving to a lattice site occupied by an
obstacle. Unlike previous work modeling lipid rafts, we closely
examine the dependence on binding, instead of just pure exclusion or
free entry into lipid regions \cite{nicolau_sources_07}.

To understand the effects of bound mobililty, we first consider the
limits of `sticky obstacles', in which tracers are immobile while
bound, and `slippery obstacles', in which tracers are mobile while
bound.  We use lattice Monte Carlo methods to explore a range binding
energy and obstacle filling fraction.  We also examine the effects of
semi-sticky obstacles---\textit{i.e.,} intermediate bound diffusion
coefficient---and obstacle size effects, which demonstrates how
diffusion is altered in a crowded environment with compartments with
different properties---such as a cell \cite{berry_monte_02}---is
altered.  Our results demonstrate how binding and bound-state motion
independently impact particle dynamics, including long-time normal
diffusion and anomalous diffusion.  Bound tracer mobility increases
the long-time diffusion coefficient, reduces the transient anomalous
time, and eliminates caging for all times typically observed above the
percolation threshold. These results demonstrate that mobility of
bound particles can benefit biological systems by allowing mobility
even in highly crowded environments.

\begin{figure}[t]
	\centering
  \includegraphics[width=0.5\textwidth]{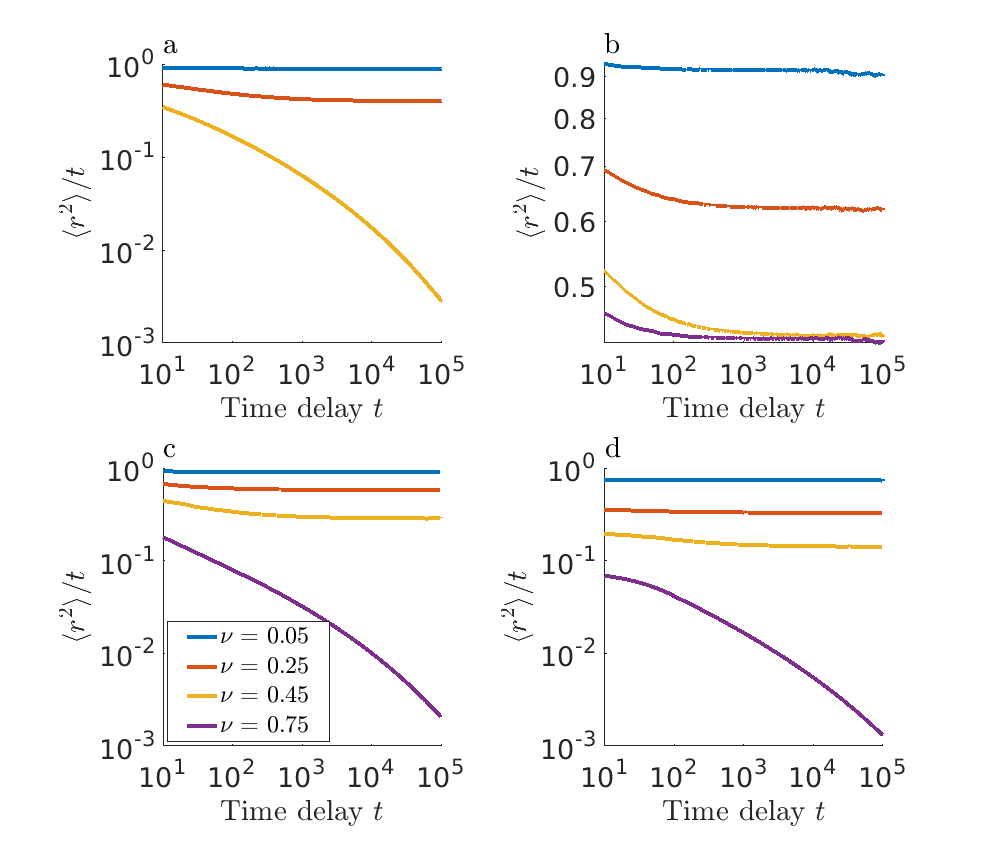}
  \caption{Mean-squared displacement $\langle r^2 \rangle$ divided by
    time delay $t$ as a function of time delay $ t $ for (a) impenetrable
    obstacles, (b) repulsive slippery obstacles ($ \Delta G = 2 $),
    (c) repulsive sticky obstacles ($ \Delta G = 2 $), and (d)
    attractive sticky obstacles ($ \Delta G = -2 $). Different colors
    correspond to different filling fraction $\nu$. Curves with
    non-zero slope indicate anomalous diffusion, and the horizontal
    asymptote indicates the long-time diffusion coefficient. Each
    curve represents an average over tracers, independent time
    windows, and obstacle configurations.}
  \label{fig:curves2fit}
\end{figure}

\section{ Model }

\begin{figure*}[t!]
	\centering
  \includegraphics[width= \textwidth]{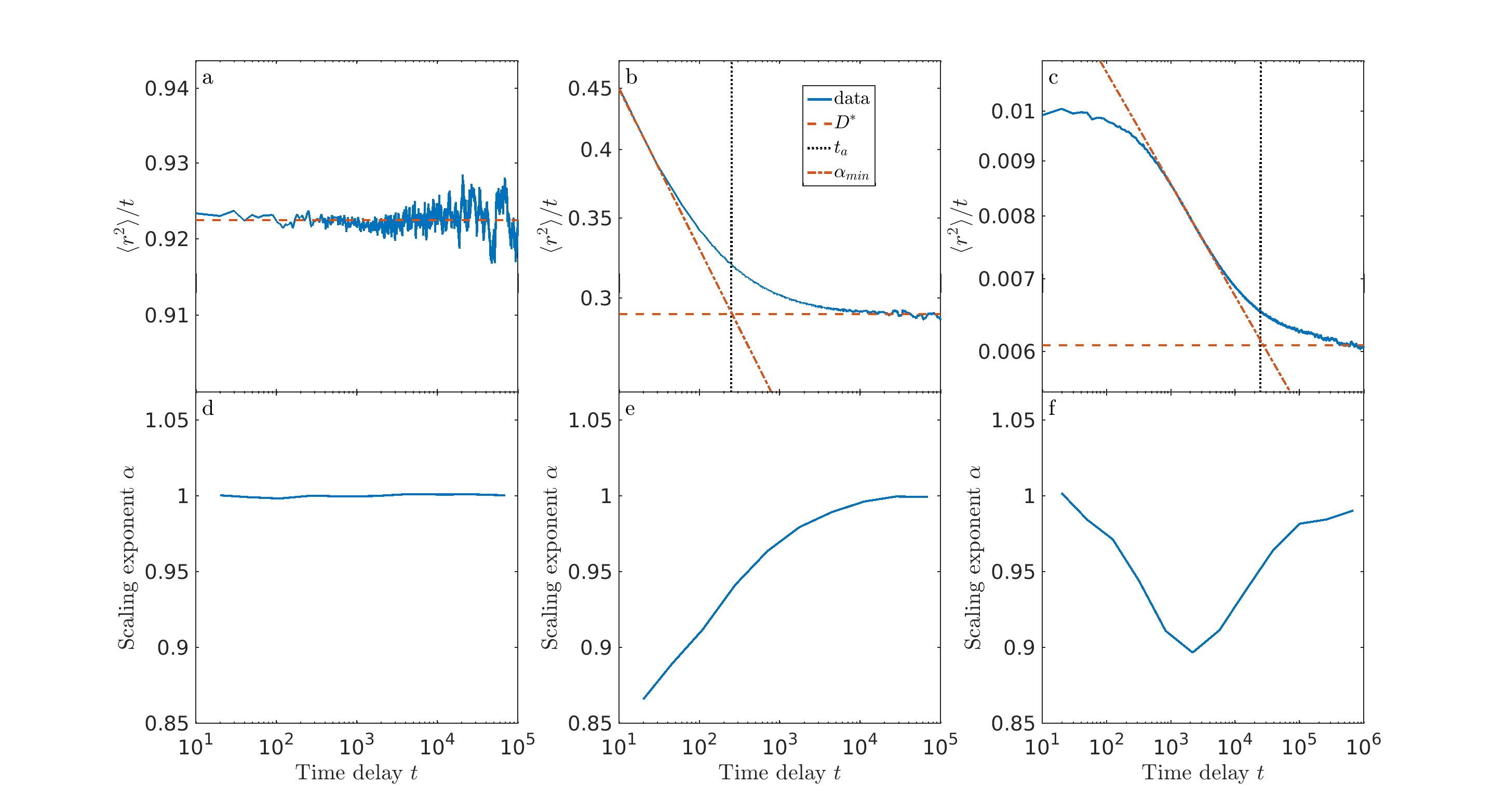}
  \caption{Top panel: Illustration of fitting procedure, showing
    $ \langle r^2\rangle /t $ vs time delay $ t $ for simulation data (blue),
    line fitted to horizontal asymptote (red dashes), line tangent to
    point of maximum absolute slope of the curve (red dash-dots), and
    anomalous time $ t_a $ (black dots) for different
    parameters. Bottom panel: instantaneous scaling exponent
    $\alpha$ vs time delay $t$. (a, d) Slippery obstacles with $ \Delta G = 1 $,
    $ \nu = 0.95 $: normal diffusion occurs for all measured time.
    (b, e) Sticky obstacles with $ \Delta G = 2 $, $ \nu = 0.45 $.
    (c, f) Sticky obstacles with $ \Delta G = -5 $, $ \nu = 0.50 $. }
  \label{fig:fittedcurve}
\end{figure*}

Our model seeks to build on stochastic lattice-gas models that have
been important to understanding tracer dynamics in the presence of
immobile and mobile hard obstacles \cite{saxton_lateral_87}, anomalous
subdiffusion \cite{saxton_anomalous_94}, and effects of binding on
diffusion \cite{saxton_anomalous_96}. Saxton showed that the tracer
diffusion coefficient drops to zero at the percolation threshold, the
critical concentration of obstacles at which a continuous path of
vacancies through which a tracer can move no longer exists. Above this
percolation threshold, diffusion is anomalous at long times.  The
effects of tracer and obstacle size
\cite{ellery_characterizing_14,ellery_calculating_15,ellery_modeling_16}
and adhesion and repulsion to sites adjacent to obstacles
\cite{ellery_analytical_16} on transient subdiffusion and long-time
diffusion have been studied. Extensions to mobile obstacles which
interact with each other have demonstrated how obstacle clustering
dynamics can influence the diffusivity of tracers
\cite{nandigrami_gradientdriven_17}. Numerically exact methods for
calculating diffusion coefficients using the Nernst-Einstein relation
\cite{mercier_numerically_99, mercier_numerically_99a} and Markov
chains \cite{ellery_communication_16} have been implemented as a
different approach to analyzing these systems; the Nernst-Einstein
approach can lower the computational cost of measuring diffusion
coefficients for lattice gases \cite{ellery_calculating_15}.  Protein
motion in polymer networks has been studied using random-walk and
self-avoiding-chain models for immobile \cite{wedemeier_modeling_07}
and mobile \cite{wedemeier_how_09, wedemeier_anomalous_09} hard
chains. Studies of chains with binding sites found that modeling chain
dynamics allowed a mapping onto randomly distributed obstacles with an
effective volume, and showed how sliding along a defined chain can
effect tracer dynamics \cite{wedemeier_role_08,
  wedemeier_anomalous_09}.  In some previous work, the effects of
binding and sliding while bound were entangled because both effects
were encoded by a single parameter \cite{wedemeier_role_08,
  wedemeier_anomalous_09}.  Domains with different diffusion
coefficients and sizes---to model lipid rafts---have been studied, but
the analysis only included total or no exclusion although it was noted
that binding effects could play a large role
\cite{nicolau_sources_07}.

In our model, tracer particles undergo a random walk on a square
lattice and interact with immobile obstacles. The interaction is
characterized by a binding free energy; for simplicity, we neglect any
additional activation barrier.  The characteristic binding free energy
change of a tracer that hops from an empty site to an obstacle site is
$ \Delta G $ (in units where $k_B T = 1$).  We consider both
attractive ($ \Delta G < 0 $) and repulsive ($ \Delta G > 0 $)
obstacles. We use the Metropolis algorithm
\cite{metropolis_equation_53} to accept or reject candidate binding
(probability $P_B$) and unbinding (probability $P_U$) events. Each
tracer occupies a single site lattice site, but the obstacle size is
varied to represent domains of characteristic size. Obstacles are
squares with sides of length $l_{\rm obst}$, measured in units of the
lattice spacing.
 
To study the effects of tracer particle motion while bound, we
considered the limits of perfectly sticky and slippery obstacles
\figref{fig:modelcartoon}, as well as the intermediate `semi-slippery'
case. In all models, obstacles are soft, so that tracers overlap with
obstacles when bound. For sticky obstacles, no hopping between
obstacle sites can occur, but tracers can exit an obstacle in any
direction that would move the tracer to an unoccupied site.  For
slippery obstacles, tracers can hop between adjoining obstacles while
remaining bound.  In the limit of perfectly slippery obstacles, in
which bound motion is identical to unbound motion, there is no
difference in hopping rates between free and bound tracers. For
semi-slippery obstacles, we vary the bound diffusion coefficient.

\subsection{ Simulation methods}
In our kinetic Monte Carlo scheme, at each time step a tracer attempts
a move in a randomly chosen direction. Moves from
$ \mathbf {empty} \rightarrow \mathbf {empty} $ are always accepted,
$ \mathbf {empty} \rightarrow \mathbf {obstacle}$ moves are accepted
with probability $ \mathbf {\rm min}( e^{-\Delta G}, 1) $,
$ \mathbf {obstacle} \rightarrow \mathbf {empty} $ moves are accepted
with probability $ \mathbf {\rm min}( e^{\Delta G}, 1 ) $, and
$ \mathbf {obstacle} \rightarrow \mathbf {obstacle} $ moves are always
accepted/rejected if obstacles are slippery/sticky
\figref{fig:modelcartoon}; for semi-slippery obstacles, the acceptance
probability is $D_{\rm bound}/D_{\rm free}$. If a tracer's move is
rejected, it remains immobile for that time step. We assume
noninteracting tracers.

Initially, obstacles were uniformly randomly placed on the lattice, at
the specified filling fraction, without overlaps. Next, tracers were
randomly placed on obstacles and empty sites at their equilibrium
occupancy, as determined by the filling fraction of obstacles $\nu$,
and binding energy $\Delta G$.  The fraction of tracers on obstacles
is proportional to the obstacle filling fraction times the Boltzmann
factor, $\nu e^{-\Delta G}$, while the fraction of tracers on empty
sites is proportional to the fraction of empty sites, $(1-\nu)$. The
equilibrium fraction of tracers on obstacles of size 1 is then
\begin{equation}
  \mathit{f}_{o} = 
    \frac{ \nu e^{-\Delta G} } {  \nu e^{-\Delta G} + \left( 1 - \nu \right) }.
\end{equation}
Using an initial fraction of tracers bound to obstacles determined
from $f_o$ avoids the time required for binding equilibration in the
simulations, ensuring that mean-squared displacement measurements are
independent of a time origin.

We performed 2D simulations with $ 200 $ tracers on a
$ 256 \times 256 $ periodic lattice for $ 10^{5} - 10^{7.5} $ time
steps, with recording interval of $ 10 - 100 $ steps. For each
parameter set (determined by filling fraction and binding energy), we
averaged over $ 96 $ separate obstacle configurations. We varied $\nu$
from 0 to 1 and $\Delta G$ from $-5$ to $10$. 3D simulations used
similar parameters with a $ 256 \times 256 \times 256$ periodic
lattice. In the semi-slippery case, we varied the ratio of bound to
free diffusion coefficient $D_{\rm bound}/D_{\rm free}$ between 0
(perfectly sticky) and 1 (perfectly slippery) in steps of 0.2, for
binding energies $\Delta G = 1, 2, 3, \infty$ for two filling
fractions, $ \nu = 0.3 $ and $ 0.6 $.  When varying obstacle size, we
used square obstacles with the length of a side, $l_{\rm obst}$, equal
to odd values from 1 to 15.

\subsection{Trajectory analysis}

We determined tracer mean-squared displacement (MSD) as a function of
time delay by averaging over all tracers, $ 100 $ randomly selected
independent time origins, and obstacle configuration. For long time
delays for which $100$ independent time intervals were not available,
we averaged over the maximum number of independent time intervals. As
previously mentioned, averaging over time windows improves our
statistics; note that the time origins are not unique, since the
placement of tracers in their equilibrium binding distribution ensures
that there is no initial binding equilibration time.  We have verified
that that are no aging effects \cite{magdziarz_fractional_09,
  schulz_aging_14}, \textit{i.e.}, MSD measurements that depends on
simulation time, in our model (data not shown).
 
We sought to quantify the effects of binding and obstacle filling
fraction on tracer mobility.  In systems with either purely Fickian
diffusion or particular obstacle geometry, the mean-squared
displacement grows as a power law in time:
\begin{equation}
  \label{eqn:msd}
  \langle r^2 \rangle =  2dD t ^ {\alpha},
\end{equation}
where $\langle r^2 \rangle$ is the ensemble-, time-origin-, and
obstacle-configuration-averaged mean-squared displacement, $d$ is the
spatial dimension, $D$ is the diffusion coefficient, $\alpha$ is the
diffusion scaling exponent, and $t$ is the time delay. This fractional
diffusion equation has been studied extensively
\cite{metzler_random_00}, both because it emerges from certain
microscopic theories and as a means to quantify anomalous random
walks.  Fractional diffusion has been experimentally measured in
cells, using fluorescence recovery after photobleaching
\cite{feder_constrained_96}, fluorescence correlation spectroscopy
\cite{weiss_anomalous_04}, and single-particle tracking
\cite{ghosh_automated_94}.  For hard obstacles, $ \alpha $ reflects
the non-homogeneity and fractal structure of a cluster. In this case,
$\alpha$ can be thought of as a measure of a local landscape, in which
obstacles have the possibility of trapping a tracer and introducing
memory effects into the system. The value of $\alpha$ does not
quantify the time it takes to escape a trapping cage; but $\alpha < 1$
suggests the possibility that the landscape can cage tracers.  In the
$\alpha \rightarrow 0$ limit, a tracer is fully caged, and the
$\alpha \rightarrow 1 $ limit represents Fickian diffusion.

However, many systems have more complex dynamics that are not power
law.  For example, tracer dynamics can be transiently anomalous:
subdiffusive on short time scales and Fickian on longer time scales
\figref[b]{fig:curves2fit}.  The dynamics can be quantified using a
phenomenological approximation in which the exponent $\alpha$ is
treated as time dependent \cite{saxton_anomalous_94,
  saxton_singleparticle_97, ellery_characterizing_14,
  ellery_calculating_15, ellery_communication_16, ellery_modeling_16}.
Thus, $ r^2 \sim t^{\alpha} $ holds only over particular time scales.

For non-power-law dynamics, we can apply equation \ref{eqn:msd}
locally, with a phenomenological, time-varying exponent. Then
$\alpha(t)$ is defined by local fitting to the the logarithm of
$\frac{ \langle r^2 \rangle } { t }$:
\begin{equation}
  \label{log_msd}
  \log \left(  \frac{ \langle r^2 \rangle } { t } \right)  
    = \log \left( 2dD  \right) + ( \alpha(t) - 1 ) \log \left( t \right).
\end{equation}
so that $\alpha(t) - 1 $ is the local slope of the
$\frac{\langle r^2 \rangle} {t}  $ versus $t$ curve on a
log-log plot.  As seen in figs.~\ref{fig:curves2fit} and
\ref{fig:fittedcurve}, the instantaneous effective $\alpha$ varies
with delay time.  Thus, a power-law MSD scaling with time, such as can
arise from fractional Brownian dynamics, does not encompass the
complexity of our crowded diffusion model, as has been found
previously \cite{ellery_modeling_16, jeon_protein_16}.

At short time, our model typically exhibits anomalous
diffusion. However, in some conditions, the short-time behavior is
diffusive, with an intermediate anomalous regime.  We defined
$\alpha_{\rm min}$ as the minimum instantaneous value of $\alpha$ (the
most anomalous exponent).  We characterized the
transition between short- or intermediate-time anomalous diffusion and
long-time normal diffusion by the time scale $t_a$, determined as the
intersection of the horizontal long-time asymptote of
$ \frac{ \langle r^2 \rangle } { t } $  with a line tangent to the point of the maximum rate of decrease of this curve \figref[b,c]{fig:fittedcurve}. We found that this transition time could be robustly determined for a wide range of
diffusion coefficients and anomalous behavior. We denote $t_{a}$ the anomalous
time. Qualitatively, it is the crossover time from short-time
subdiffusion to long-time Fickian diffusion. While $\alpha_{\rm min}$
characterizes how trapped a tracer is, $t_{a}$ quantifies how long it
takes a tracer to escape a cage.

\begin{figure*}[t]
	\centering
  \includegraphics[width=1.0\textwidth]{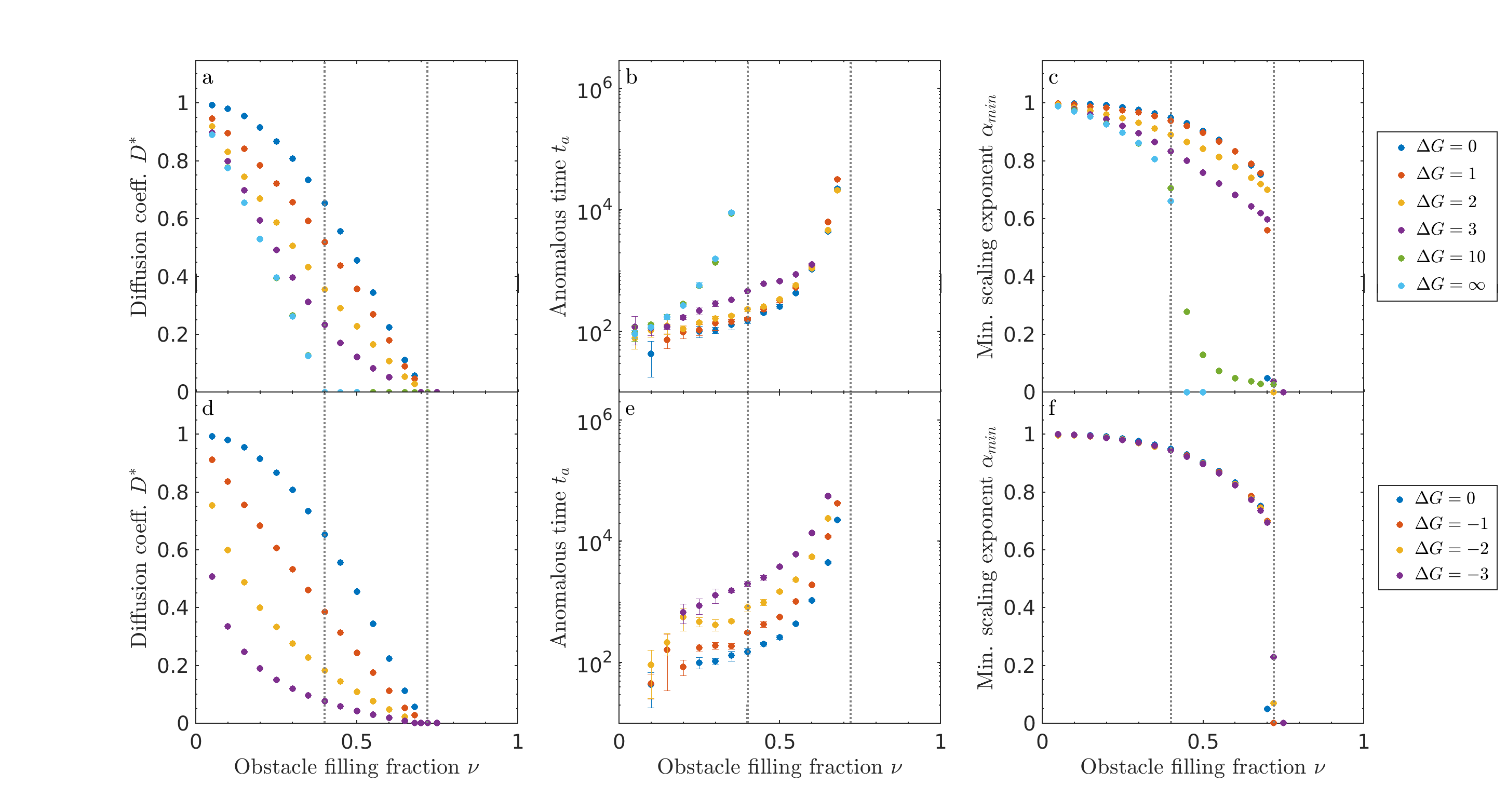}
  \caption{Sticky obstacles of size 1 in 2D.  (a, d) Diffusion
    coefficient $D^*$, (b, e) anomalous time $ t_a $, and (c, f)
    minimum scaling exponent $\alpha_{\rm min}$ as a function of
    obstacle filling fraction $\nu$ for positive (top) and attractive
    (lower) binding energy. Note that points for $\Delta G = 10 $ are
    partially hidden behind $ \Delta G = \infty$.  The approximate
    locations of the critical occupancies $ \nu^l $ and $ \nu^u $ are
    indicated with gray dotted lines.}
  \label{fig:2d_sticky}
\end{figure*}

We defined the long-time Fickian diffusion coefficient as
\begin{equation}
  \label{diff}
  D = \lim_{t \to \infty} \frac{ \langle r^2 \rangle } { 2d t }.
\end{equation}
All diffusion coefficient measurements are expressed in terms of the
scaled diffusion coefficient $ D^* = \frac{ D } {D_0} $, where
$ D_0 = \frac{ l^2 }{ 2 d \tau} $ is the diffusion coefficient in the
absence of obstacles, where $l$ is the distance between lattice sites
(here defined to be 1), and $\tau$ the time interval between steps
(also set to 1).

In some cases, we were unable to determine all of $D^*$,
$\alpha_{\rm min}$, and $t_a$. For some parameter sets, the slope of
$\frac{r^2}{t} $ vs.  $ t $ on a log-log plot approached a non-zero
constant, indicating that diffusion was anomalous over all measured time
delays, so that the Fickian diffusion coefficient was not well-defined. For
other parameter sets, the $\frac{r^2}{t} $ versus $ t $ curve did not
reached a clear asymptote during the simulation time. We therefore could
not determine $D^*$, but could measure $\alpha_{\rm min}$.  When
tracer diffusion was normal over all or nearly all measured time
delays, neither $\alpha_{\rm min}$ nor $t_a$ were well-defined, but
$D^*$ could be measured.

\begin{figure}[b!]
  \centering
  \includegraphics[width=0.5\textwidth]{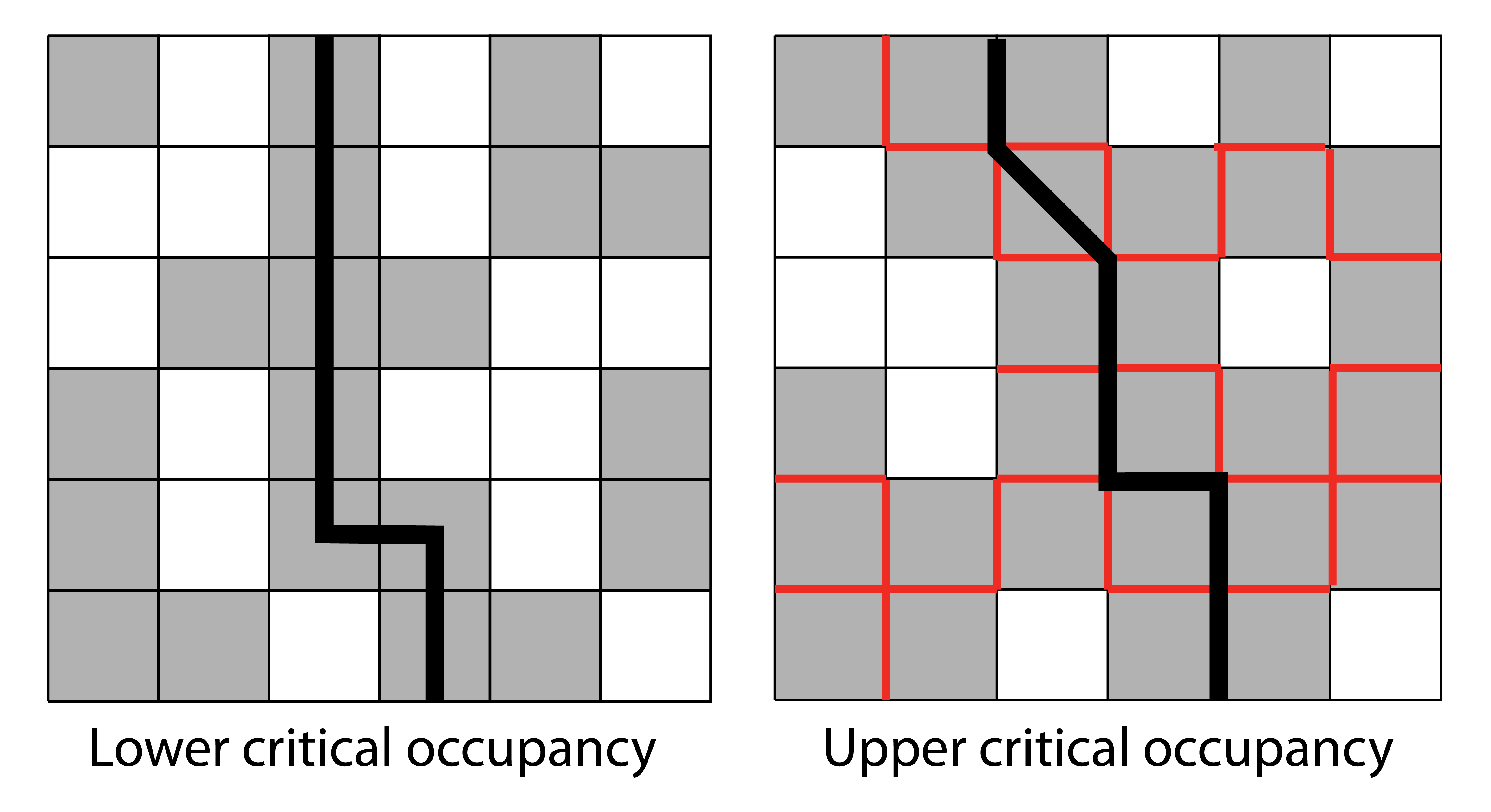}
  \caption{ The two types of percolation threshold in our lattice
    model: the lower critical occupancy $ \nu^l $ (left) and the upper
    critical occupancy $ \nu^u $ (right). For the lower critical
    occupancy, which is the standard percolation threshold, the
    percolating network is the obstacles. At the upper critical
    occupancy, the percolating network is the \textit{interface}
    between two or more obstacles. The barrier to tracer motion is
    shown as a black lines; obstacle-obstacle boundaries which cannot
    be crossed by a tracer in the sticky model are shown in
    red. Without binding, tracers cannot pass through the lower
    percolating network. If they can bind, tracers can `hop through'
    the lower percolation barrier with or without bound motion.}
     \label{fig:percGrid}
\end{figure}

\section{Sticky soft obstacles}

We initially focus on the limit of perfectly sticky obstacles of size
1, to determine the effects of stickiness, filling fraction, and
binding energy on tracer motion. We varied parameters over a wider
range for the 2D model, with a comparison to 3D results for some
parameter sets.

For sticky obstacles, the motion of a bound tracer to an adjacent obstacle
is prohibited.  This could occur, for example, because the net free
energy cost of binding to an obstacle is a result of an attractive
binding interaction, with a high free energy barrier to moving to
an adjacent site.  Here, we consider the
limit that the free energy cost of moving to an adjacent obstacle is
so large that it approaches infinity.  This situation provides an
important point of comparison to explicitly test the effects of
bound-state diffusion on tracer behavior.

We separately consider repulsive and attractive obstacles
\figref{fig:2d_sticky}.  Note that we include the case $\Delta G=0$,
that is, where the binding interactions are neither attractive nor
repulsive, but still block moves to adjacent obstacles.  We define the
lower critical occupancy $\nu^l$ as the filling fraction at which the
diffusion is non-Fickian for all time scales for impenetrable
obstacles ($\Delta G = \infty$).  In the limit of a hard repulsive
obstacle, $D^*$ decreases with filling fraction, and approaches zero
at the percolation threshold expected for hard obstacles on a square
lattice, $\nu^l \approx 0.4$ \cite{stauffer_introduction_94}, where
$t_a$ diverges \cite{saxton_lateral_87}.  The lower critical occupancy
is the percolation threshold, at which there is no longer a continuous
path of empty sites \figref{fig:percGrid}.

\begin{figure*}[t]
	\centering
  \includegraphics[width=1.0\textwidth]{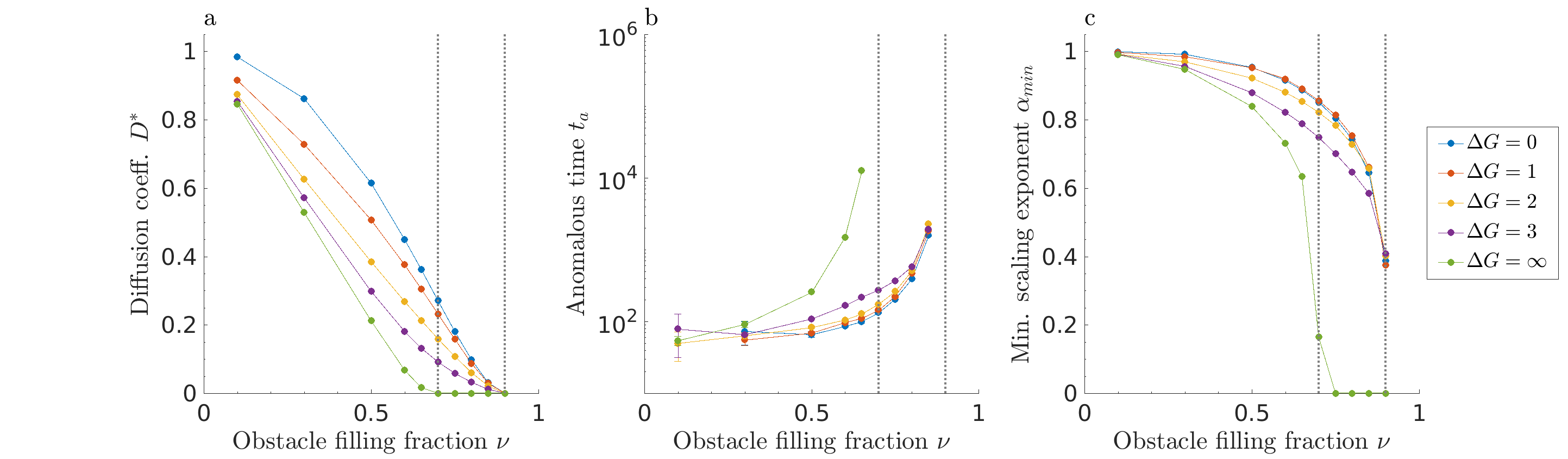}
  \caption{ Sticky obstacles of size one in 3D. (a) Diffusion
    coefficient $D^*$, (b) anomalous time $ t_a $, and (c) minimum
    scaling exponent $\alpha_{\rm min}$ as a function of obstacle
    filling fraction. The approximate locations of the critical
    occupancies $ \nu^l $ and $ \nu^u $ are indicated with gray dotted
    lines.}
  \label{fig:3d_sticky}
\end{figure*}

For finite binding free energy in our model, Fickian diffusion can
still occur above the percolation threshold $\nu^l$ because soft
binding allows tracers to `hop through' single obstacles via binding
and unbinding. Without soft binding of the type we consider, obstacle
percolation would prevent a tracer from moving between vacancy
clusters. In other words, tracers that start in an area caged by
obstacles are stuck there.  With soft binding, tracers that start in a
cage can hop onto an obstacle and then hop off into a new vacancy
cluster.  For soft binding interactions and sticky obstacles, there is
an upper critical occupancy $ \nu^u \approx 0.72 $ at which the
long-time diffusion coefficient approaches zero irrespective of
binding energy \figref{fig:2d_sticky}. Above $\nu^u$, tracers become
caged regardless of the binding kinetics.  Therefore, there is a
different type of percolating network above the upper critical
occupancy: the percolation of the inter-obstacle boundary
\figref{fig:percGrid}.  At the upper critical occupancy, there is a
second adjacent obstacle preventing the tracer from `hopping through.'
Note that as expected, the transition time $t_a$ appears to diverge on
the approach to the upper critical occupancy \figref{fig:2d_sticky}.
We are unaware of a theoretical value for this percolating density,
but our results suggest its approximate value is $0.72$ in 2D
\figref{fig:2d_sticky}.

Intermediate repulsive binding energy leads to intermediate behavior,
as expected. For strong repulsion, \textit{e.g.}, $\Delta G=5$, $D^*$
remains small, though clearly non-zero, up to the upper critical
occupancy, while $t_a$ monotonically increases until it diverges at
$\nu^u$.

Anomalous dynamics appear in the slope of $ \langle r^2 \rangle / t $
on a log-log plot. The most anomalous behavior occurs when the scaling
coefficient $ \alpha $ reaches its smallest value, $\alpha_{\rm min}$.
We find that $\alpha_{\rm min}$ decreases with filling fraction and
binding energy \figref[c]{fig:2d_sticky}. Adding more obstacles and
increasing the repulsion causes greater hindrance of tracer motion.
We note that $\alpha_{\rm min} \approx 0.7$ near $\nu^l$ for
impenetrable obstacles, as found previously \cite{saxton_anomalous_94,
  nicolau_sources_07}.  Finite repulsive binding energy leads to a
smaller exponent ($ \alpha_{\rm min} < 0.7 $) than the infinite case
at filling fraction above $\nu^l$. For lower values of $\Delta G$, the
scaling coefficient does not go to zero at the upper critical
occupancy $ \nu^u $.  Note that the sharp cutoff with filling fraction
occurs because we did not collect data past $ \nu^u $.

Sticky obstacles with attractive binding interactions show a more
rapid falloff in the diffusion coefficient and larger anomalous time
\figref{fig:2d_sticky}.  The upper critical density
$ \nu^u \approx 0.72 $ is in the same vicinity as for $\Delta G > 0$.
The dependencies of the diffusion coefficient on filling fraction for
positive and negative binding energy are similar for low magnitude of
the binding energy, but the diffusion coefficient falls off more
rapidly with filling fraction for highly attractive obstacles. This
occurs because an attractive obstacle confines a tracer in one
position until it escapes, while a repulsive obstacle only impedes
tracer motion for one time step. Therefore, repulsive interactions
require several obstacles to transiently confine a tracer, while a
single attractive obstacle can cause confinement. Note that we did not
include large attractive binding free energy in our analysis.

For attractive obstacles, $ \alpha_{\rm min} $ is independent of
binding energy over the range we studied \figref{fig:2d_sticky}. The
characteristic time for a tracer to unbind from an attractive obstacle
depends on the binding energy, leading to the energy-dependent
variation in the anomalous time we observe. However, it is properties
of the obstacle arrangement, rather than of binding, which determine
the shape of the MSD curve, and therefore the $\alpha_{\rm min}$.  The
minimum anomalous exponent occurs when tracers are, on average,
confined to a cage formed by inter-obstacle boundaries and single-site
wells.  Therefore, the minimum anomalous exponent is approximately the
same for all binding energy, but varies with filling fraction.

\begin{figure*}[t]
	\centering
  \includegraphics[width=1.0\textwidth]{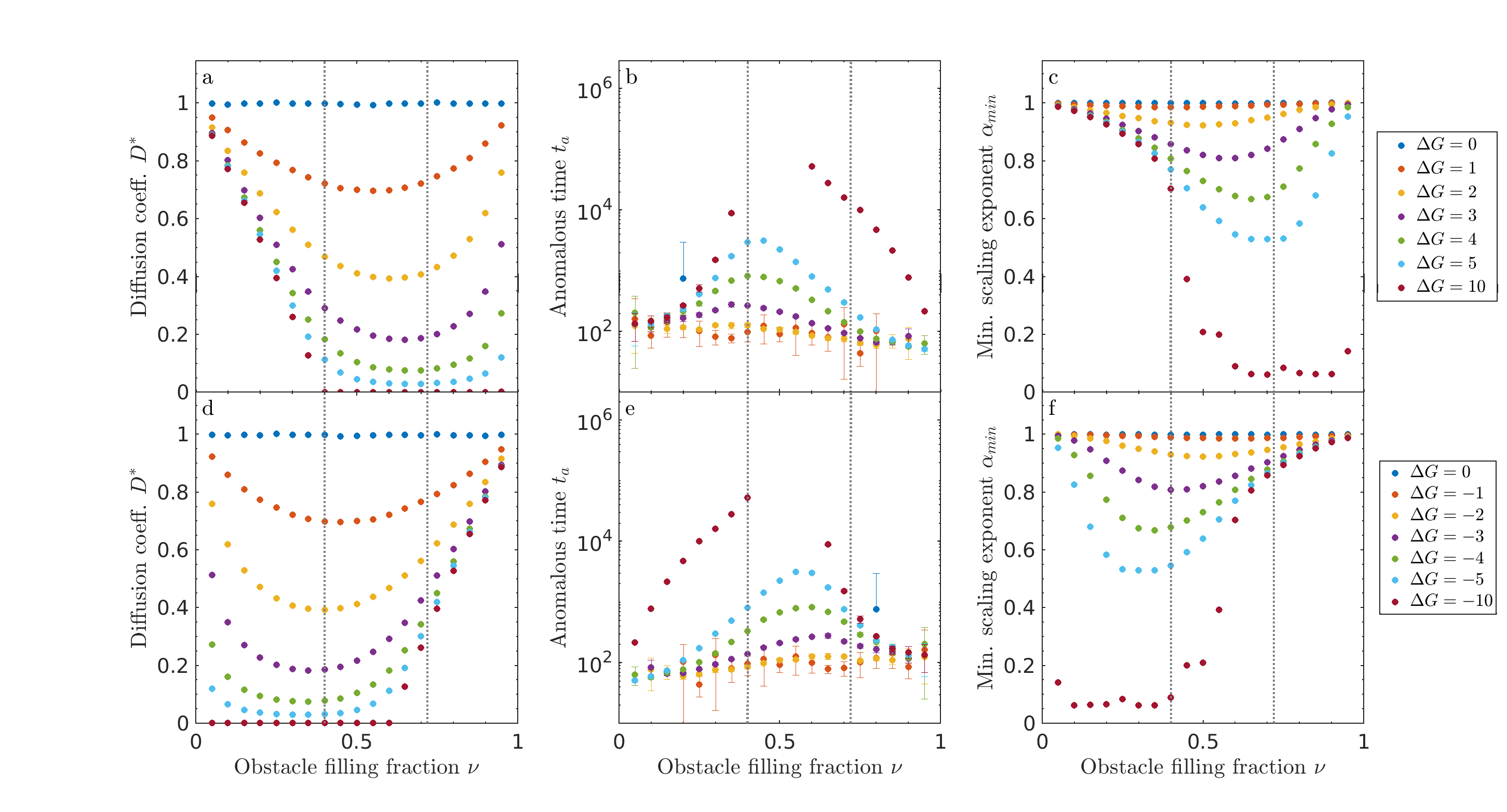}
  \caption{Slippery obstacles in 2D. (a, d) Diffusion coefficient
    $D^*$, (b, e) anomalous time $ t_a $, and (c, f) minimum scaling
    exponent $\alpha_{\rm min}$ as a function of obstacle filling
    fraction $\nu$ for positive (top) and attractive (lower) binding
    energy.  The approximate locations of the critical occupancies
    $ \nu^l $ and $ \nu^u $ are indicated with gray dotted lines.}
  \label{fig:2d_slippery}
\end{figure*}

\begin{figure*}[t]
	\centering
  \includegraphics[width=1.0\textwidth]{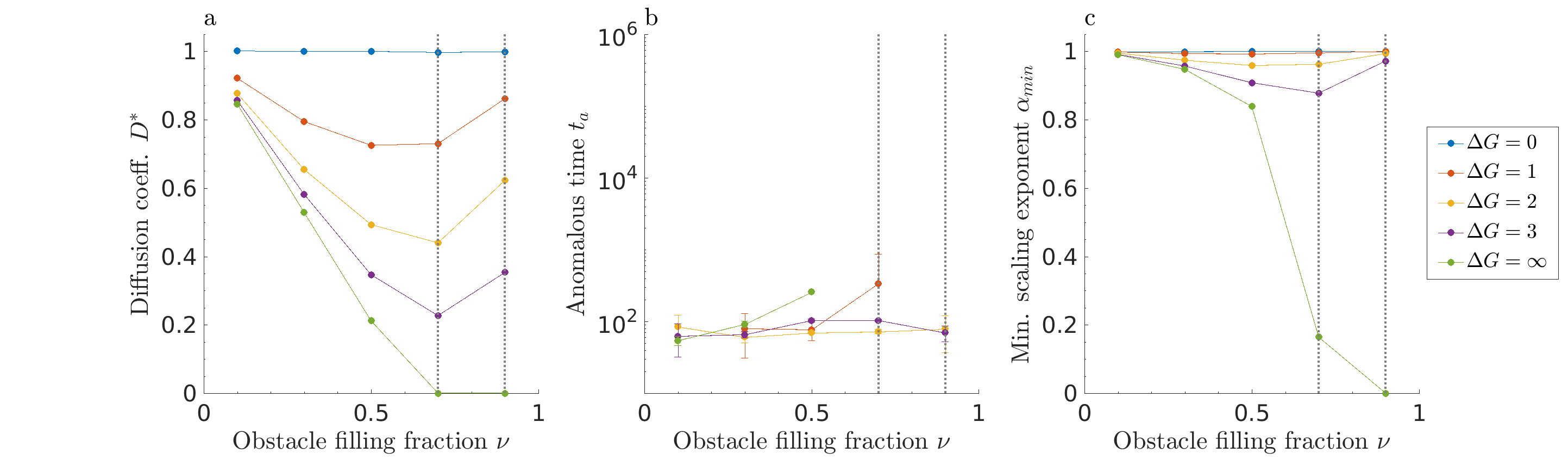}
  \caption{ Slippery obstacles of size 1 in 3D. (a) Diffusion
    coefficient $D^*$, (b) anomalous time $ t_a $, and (c)
    minimum scaling exponent $\alpha_{\rm min}$ as a function of
    obstacle filling fraction $\nu$. The approximate locations of the critical
    occupancies $ \nu^l $ and $ \nu^u $ are indicated with gray dotted
    lines. }
  \label{fig:3d_slippery}
\end{figure*}

We note that the sticky soft obstacle model studied here does not
simply map to the impenetrable obstacles at a lower effective obstacle
filling fraction. Such a mapping cannot be made because tracers can
`hop through' single obstacles via binding, while never being able to
hop between obstacles.  Sticky obstacles allow for move attempts---and
blocks---that would never be attempted in the impenetrable case.

\subsection{Sticky soft obstacles in 3D}

We extended our study of single-site sticky repulsive obstacles to
three dimensions, to determine whether the spatial dimension plays a
key role in the tracer behavior \figref{fig:3d_sticky}. The results
are qualitatively the same as the 2D model \figref{fig:2d_sticky}.
However, in 3D, the lower and upper critical occupancies appear at
higher filling fraction: a higher obstacle filling fraction is
required to percolate a 3D lattice. The anomalous time is also
typically smaller in 3D. For soft sticky obstacles, increasing the
spatial dimension does not change the qualitative features of our
model, but does shift the critical occupancies and anomalous time.

\section{ Slippery soft obstacles}

When obstacles are perfectly slippery, bound tracers can hop to
adjacent obstacles without penalty.  Our model of perfectly slippery
obstacles contains an occupancy-energy inversion symmetry: the
dynamics are invariant to changing the filling fraction by switching
obstacles and empty sites ($ \nu \rightarrow 1-\nu $) while
simultaneously switching the sign of the binding energy
($ \Delta G \rightarrow - \Delta G $). In other words, a low filling
fraction of attractive obstacles is equivalent to a high filling
fraction of repulsive barriers \figref{fig:2d_slippery}.

Slippery obstacles remove the obstacle percolation threshold for all
measured binding energies \figref{fig:2d_slippery}. The curves for
$\Delta G = 10$ for the repulsive slippery obstacles qualitatively
resemble the sticky case \figref{fig:2d_sticky}, because the diffusion
coefficient approaches zero for $\nu \approx 0.4 $. However, for
slippery obstacles, the anomalous time increases, but does not
diverge, at the percolation threshold, and then decreases at larger
filling fraction. For slippery obstacles with finite $\Delta G$, one
can always find a time after which the system displays normal
diffusion.
\begin{figure*}[t]
	\centering
  \includegraphics[width=1.0\textwidth]{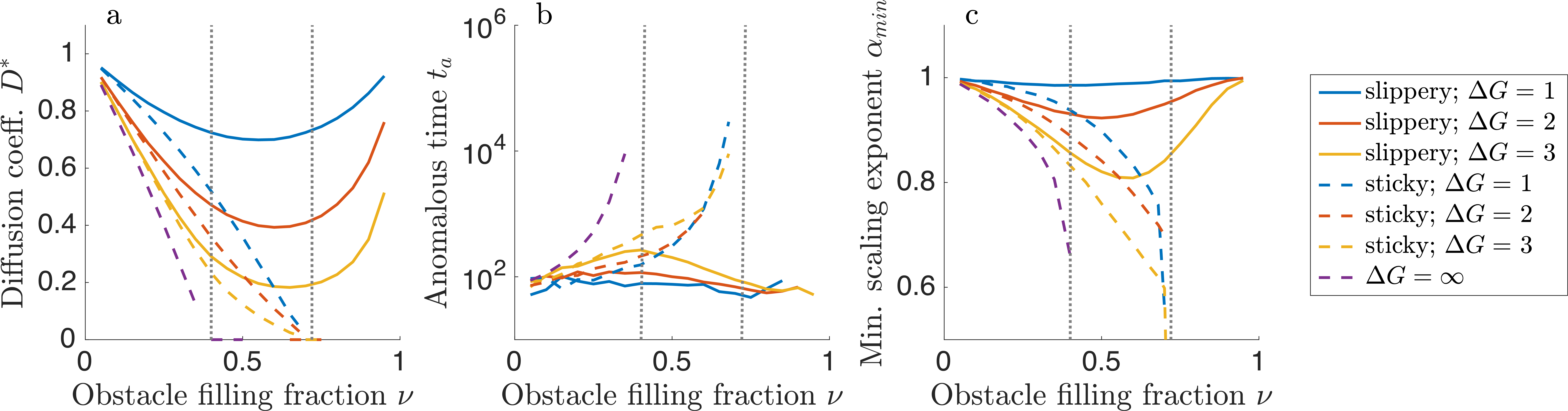}
  \caption{Comparison of models with slippery repulsive obstacles
    (solid lines), sticky repulsive obstacles (dashed lines), and hard
    repulsive obstacles (purple dashed line). (a) Diffusion
    coefficient $D^*$, (b) anomalous time $ t_a $, and (c) minimum
    scaling exponent $\alpha_{\rm min}$ as a function of obstacle filling
    fraction $\nu$.  The gray dotted lines are indicating the
    approximate locations of critical occupancies $ \nu^l $ and
    $ \nu^u $. }
  \label{fig:comprDT}
\end{figure*}
Slippery obstacles lead to non-monotonic behavior: for large enough
$\nu$, the diffusion coefficient increases and anomalous time
decreases. For high obstacle filling fraction, binding increases
tracer mobility, because they can hop along the percolating network of
obstacles.  Similarly, the minimum exponent is non-monotonic with
filling fraction. 

\subsection{Slippery soft obstacles in 3D}

As for sticky obstacles, we examined tracer motion with single-site
slippery obstacles in three dimensions \figref{fig:3d_slippery}. The
results are qualitatively the same as the 2D model
\figref{fig:2d_slippery}, with typically smaller anomalous time.

\subsection{Comparison of sticky and slippery obstacles in 2D}
The limits of perfectly sticky and slippery obstacles are most similar
at low filling fraction \figref{fig:comprDT}. In general, slippery
obstacles lead to exponents closer to one (less anomalous) than do
sticky obstacles, because tracers are not caged by the
obstacle-obstacle interface.  Even for relatively small values of the
binding energy ($|\Delta G| \le 3$) and intermediate filling fraction,
sticky and slippery obstacles lead to significantly different tracer
dynamics \figref{fig:comprDT}. Slippery obstacles, on which motion can
occur for high obstacle filling fraction, allow normal diffusion with
coefficients comparable to those for low filling fraction. This effect
may be important to explain the rates of a number of biological
processes that are diffusion-limited, including transcriptional
regulation and nucleo-cytoplasmic transport.

\section{ Semi-slippery obstacles}
\begin{figure*}[t]
	\centering
  \includegraphics[width=1.0\textwidth]{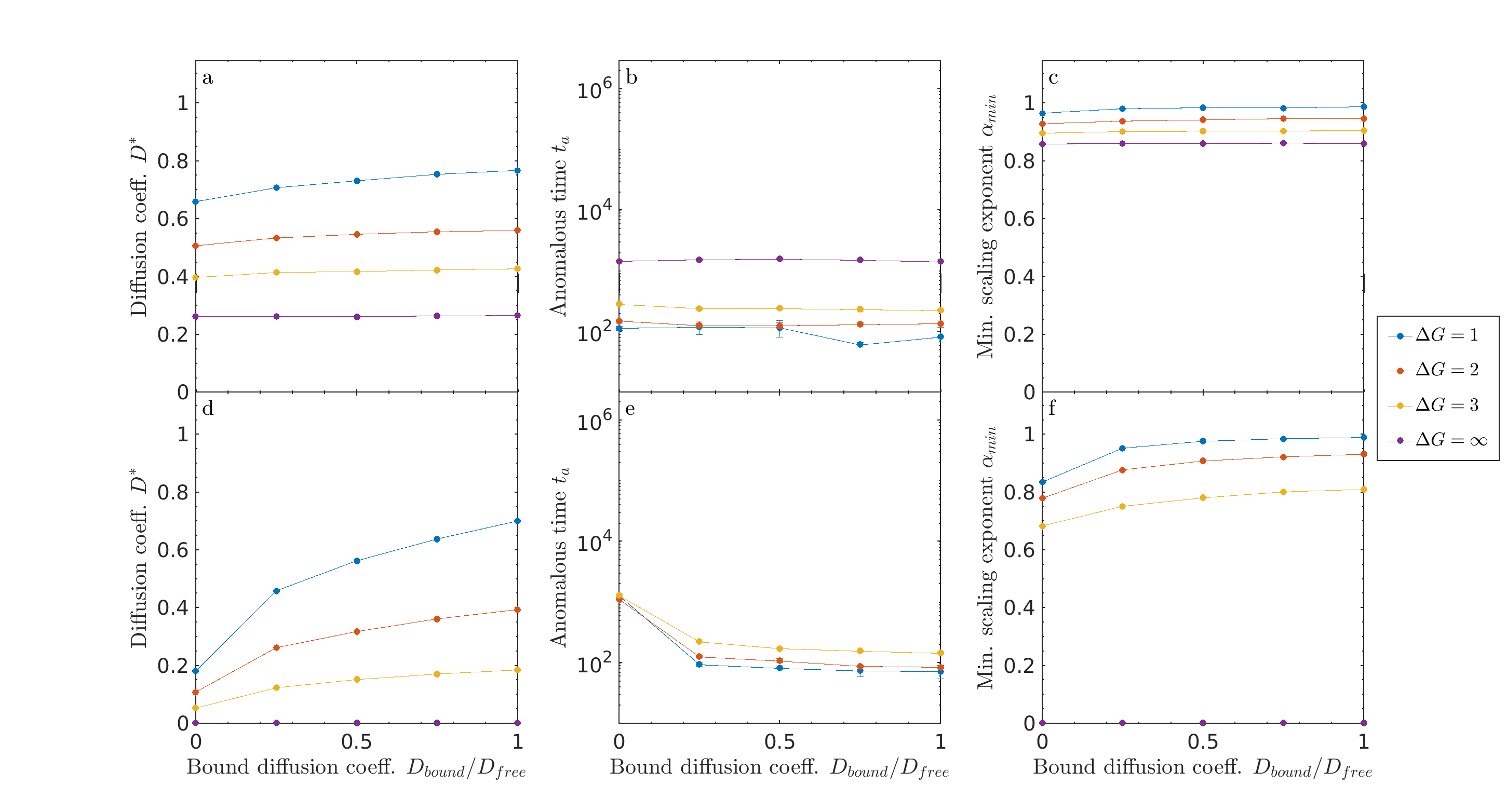}
  \caption{ Semi-slippery obstacles in 2D.  Varied the bound diffusion coefficient from the sticky $D_{\rm bound}=0$ to slippery $D_{\rm bound}=D_{\rm free}$ limit for single site $l_{\rm obst}=1$ obstacles.  Top panel: (a, d) Diffusion coefficient $D^*$, (b, e) anomalous time $ t_a $, and (c, f) minimum scaling exponent $\alpha_{\rm min}$  as function of $D_{\rm bound}$ for  low filling fraction $\nu=0.3$ (top) and high filling fraction $\nu =0.6$ (bottom).}
  \label{fig:bdiff}
\end{figure*}

Having compared the limits of perfectly sticky ($D_{\rm bound} = 0$)
and slippery ($D_{\rm bound} = D_{\rm free}$) obstacles, we now study
intermediate cases.  We varied the bound diffusion coefficient for
repulsive binding energy $\Delta G = 1, 2, 3, \infty$ and filling
fraction $ \nu = 0.3 $ and $ 0.6 $. For finite binding energy,
increasing $D_{\rm bound}$ increases the long-time diffusion
coefficient \figref{fig:bdiff}. This effect is larger for higher
filling fraction and lower binding energy, when tracers spend more
time bound.  Varying $D_{\rm bound}$ has little effect on the
anomalous time at low filling fraction, because $t_{a}$ is already
near the threshold at which we can accurately measure it. However,
increasing $D_{\rm bound}$ decreases $t_{a}$ at higher filling
fraction, because tracers can more quickly escape obstacles when their
bound diffusion coefficient is larger. Similarly, varying
$ D_{\rm bound} $ has little effect on $\alpha_{\rm min}$ at low
$\nu$, but does make diffusion less anomalous at higher filling
fraction, because increasing bound motility reduces tracer caging.

\section{Varying obstacle size}
\begin{figure*}[t]
	\centering
        \includegraphics[width=1.0\textwidth]{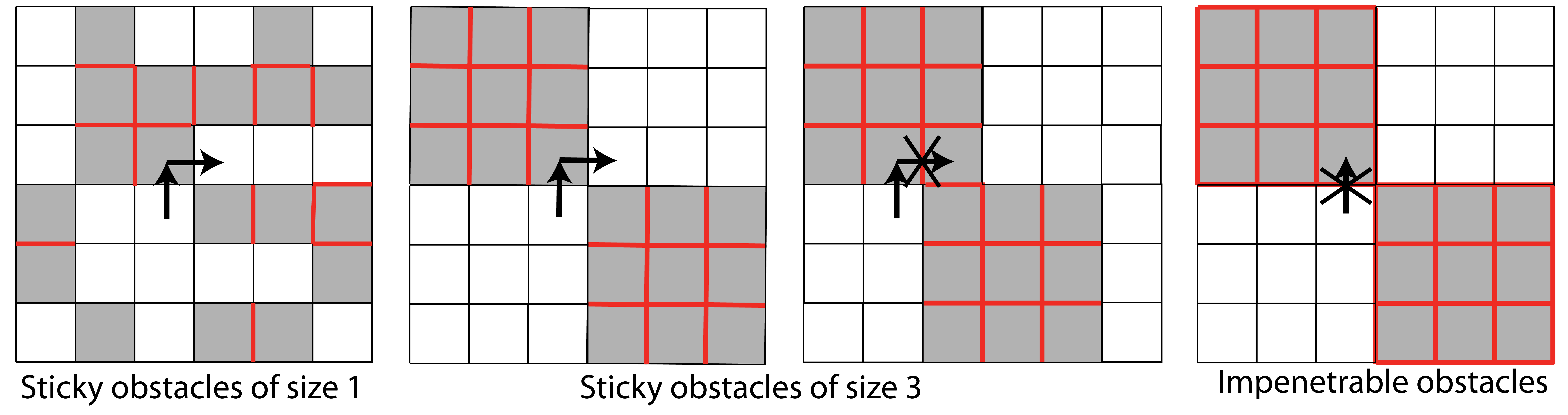}
  \caption{Cartoon showing size effects for sticky and impenetrable
    obstacles. Red lines indicate borders between obstacles that
    cannot be crossed by a tracer.}
  \label{fig:sizeCartoon}
\end{figure*}
We varied the length of the obstacles $l_{\rm obst}$, while
maintaining their square shape.  Increasing the obstacle size (with
filling fraction fixed) clusters obstacles.  Since in our model the
binding penalty occurs only for
$ \mathbf {empty} \rightarrow \mathbf {obstacle} $ moves, increasing
the size of obstacles effectively reduces the number of binding sites:
more obstacle sites are interior to obstacles, rather than on their
perimeter.  For sticky obstacles with $l_{obst} = 1$, tracers can
easily hop through cages, since their bound motion is only blocked by
an obstacle-obstacle interface.  Increasing the obstacle size
guarantees that individual obstacles will contain an obstacle-obstacle
interface, which makes it less likely that tracers can hop through
neighboring obstacles \figref{fig:sizeCartoon}.  Increasing obstacle
size at fixed filling fraction also increases the typical distance
between obstacles. These changes alter obstacle percolation effects:
$\nu^l$ and $\nu^u$ depend on $l_{\rm obst}$.

\subsection{Sticky obstacles of varying size}
\begin{figure*}[t]
	\centering
  \includegraphics[width=1.0\textwidth]{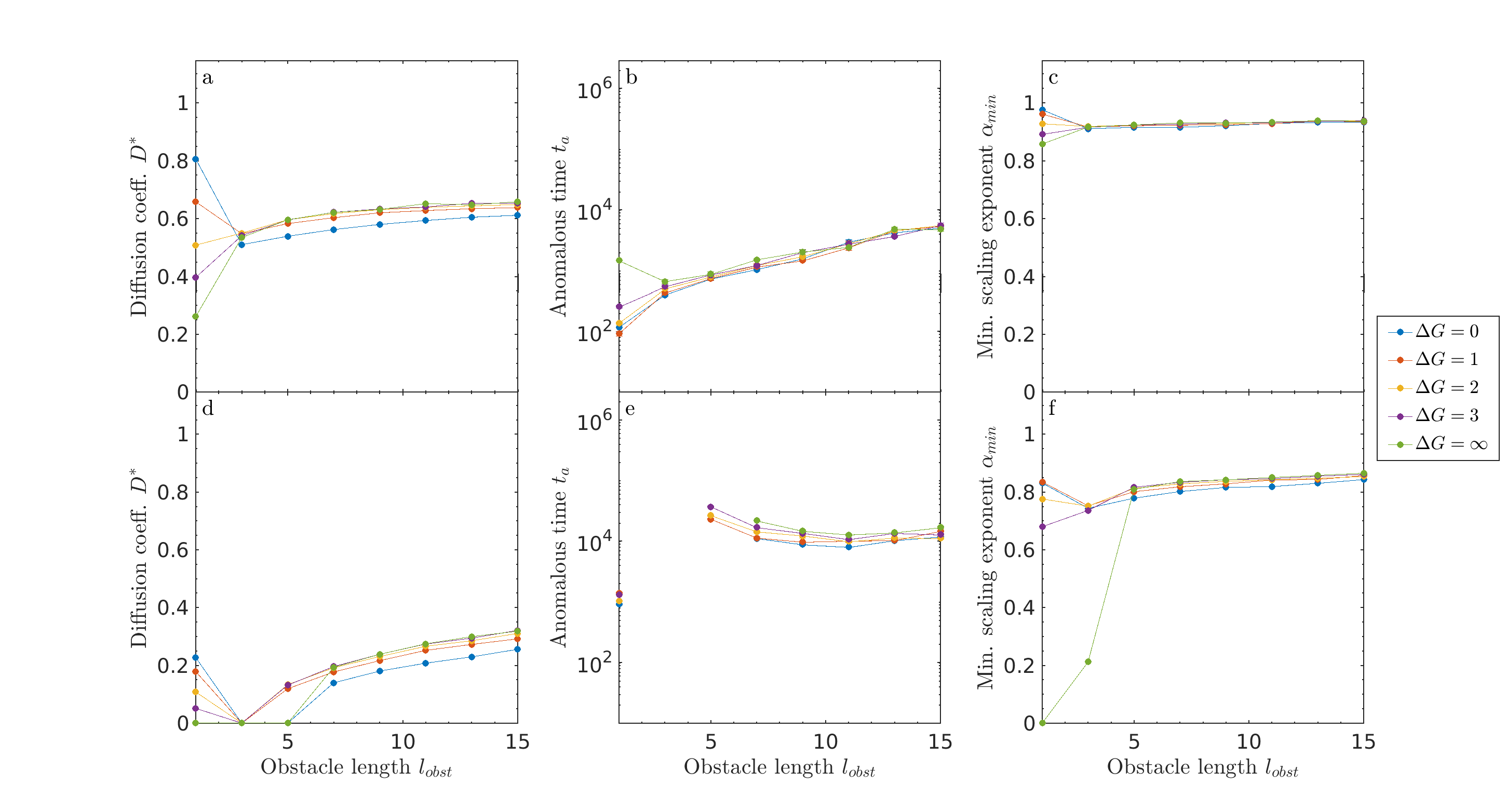}
  \caption{ Size effects for sticky obstacles in 2D.  (a, d) Diffusion
    coefficient $D^*$, (b, e) anomalous time $ t_a $, and (c, f)
    minimum scaling exponent $\alpha_{\rm min}$ as a function of
    obstacle filling fraction $\nu$ for $\nu=0.3$ (top) and $\nu =0.6$
    (lower).  }
  \label{fig:size_sticky}
\end{figure*}

First, we examined tracer dynamics on sticky obstacles of variable
size \figref{fig:size_sticky}.  Qualitatively, large sticky obstacles
have a soft surface (binding can occur on surface sites, although hops
along the surface are still blocked), but a hard core (interior sites
are inaccessible).  A significant change in dynamics occur when
$ l_{\rm obst}$ increases above 1. Any obstacle with
$l_{\rm obst} > 1$ is fundamentally different from $l_{\rm obst} = 1$,
because larger obstacles are guaranteed to contain sites with an
adjacent obstacle site. Increasing $l_{\rm obst}$ prevents hopping
across the interior of any one obstacle, which can hinder tracer
motion. The cages are thus more robust.  Tracers can still hop across
corners, unlike in the case of a purely repulsive interaction
\figref{fig:sizeCartoon}.

The dependence of tracer dynamics on binding energy changes upon
increasing the obstacle size above 1 \figref{fig:size_sticky}.  For
size-one obstacles, particles can hop through a single obstacle, and
so lower binding energy leads to higher long-time diffusion
coefficient. In contrast, with larger obstacles, high binding energy
leads to an increased diffusion coefficient. With higher repulsion, a
tracer is less likely to bind to the surface of an obstacle where it
can get stuck. Thus, for larger obstacles, higher repulsion can
facilitate motion.
 
For $l_{\rm obst}>3$, increasing obstacle size increases the cage
size, and so the long-term diffusion coefficient and the anomalous
time both increase smoothly, in agreement with previous work on
impenetrable obstacles \cite{ellery_characterizing_14}.  The anomalous
time increases with $l_{\rm obst}$ above 3, because the effective cage
size increases: tracers take longer to explore a cage to escape.  For
$l_{\rm obst}\geq 3$ and small filling fraction, the size dependence
is roughly energy independent.  The dynamics are dominated by blocked
$ \mathbf {obstacle} \rightarrow \mathbf {obstacle} $ moves, rather
than by the energy dependence of
$ \mathbf {empty} \rightarrow \mathbf {obstacle} $ moves.  For low
filling fraction, $\alpha_{\rm min}$ remains $>0.9$, suggesting that
obstacle caging effects are minimal.

Next, we examined a higher packing fraction $\nu = 0.6$, chosen
because it is between $\nu^l$ and $\nu^u$ for size-1 obstacles in
2D. The effects of obstacle size on percolation are significant,
leading to larger changes in behavior than for $\nu=0.3$.  As
$l_{\rm obst}$ increases, obstacles are on average spaced farther
apart, which increases $\nu^l$.

In contrast, the upper critical concentration is more complicated,
because now each obstacle contains within it obstacle-obstacle
interfaces.  The upper critical concentration decreases below $0.6$
for $l_{\rm obst} = 3$, and therefore the dynamics are anomalous at
all times; $t_a$ diverges and $D^*$ goes to zero.  Above
$l_{\rm obst} = 3$, the upper critical concentration increases with
increasing obstacle size.  For $l_{\rm obst} = 5$, $\nu^u>0.6$,
leading to long-time Fickian diffusion.  Here $t_a$ decreases with
$l_{\rm obst}$, because the time required for a tracer escape a cage
is not dominated by the cage size (as it was for low $\nu$), but by
the time needed to find a gap between cages.  As $l_{\rm obst}$
increases, the gaps become larger on average, lowering the escape
time.  Overall, above $l_{\rm obst} = 5$, the behavior is only mildly
dependent on either obstacle size or binding energy, making the long
term diffusivities primarily a function of the filling fraction.

\begin{figure*}[t]
	\centering
  \includegraphics[width=1.0\textwidth]{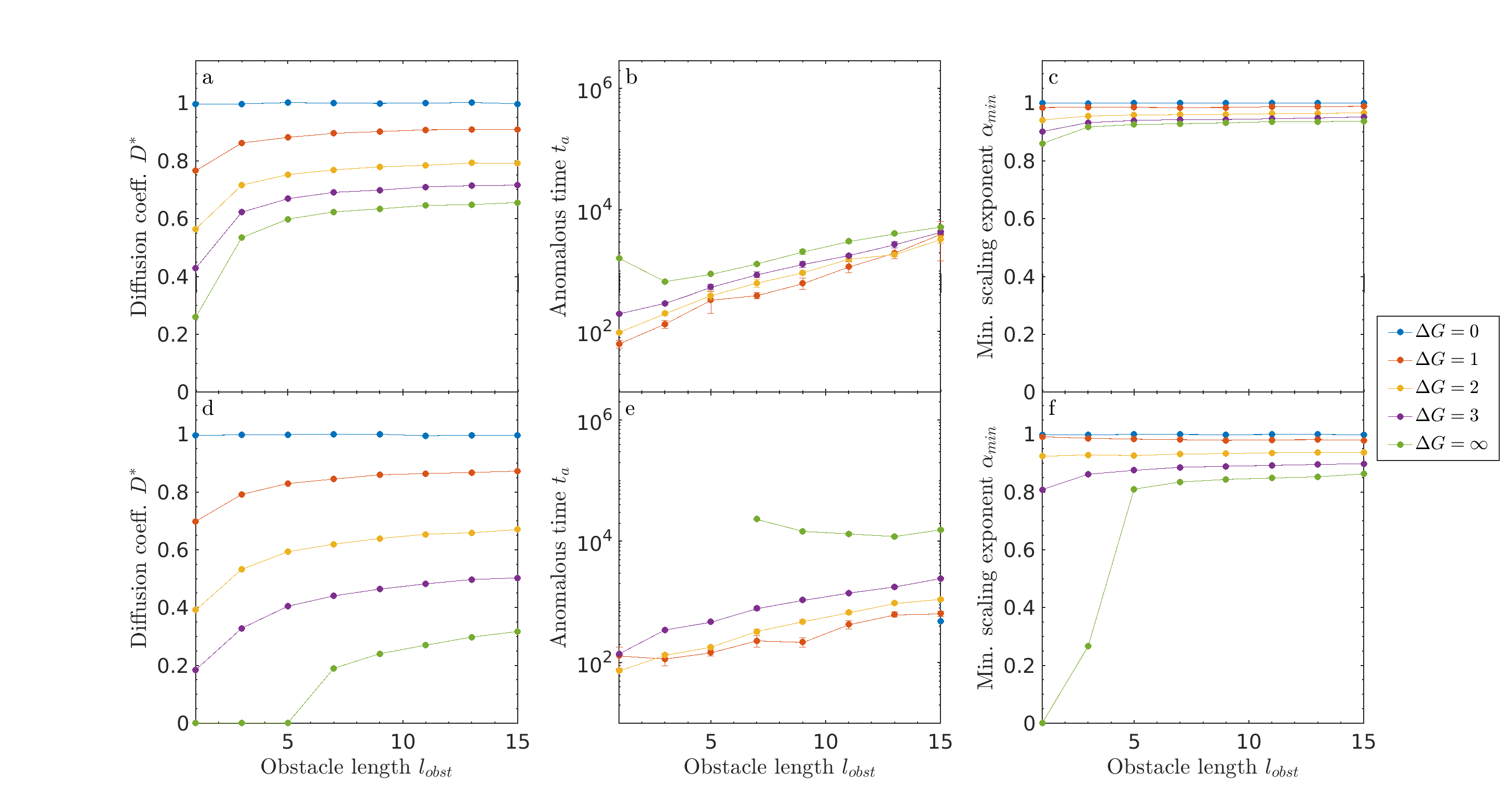}
  \caption{ Size effects for slippery obstacles in 2D.  (a, d)
    Diffusion coefficient $D^*$, (b, e) anomalous time $ t_a $, and
    (c, f) minimum scaling exponent $\alpha_{\rm min}$ as a function
    of obstacle filling fraction $\nu$ for $\nu=0.3$ (top) and
    $\nu =0.6$ (lower).  }
  \label{fig:size_slippery}
\end{figure*}

\subsection{Slippery obstacles of varying size}

Understanding the effects of variable obstacle size on tracer motion
is more straightforward for the case of slippery obstacles, because
the difference between edge and interior obstacle sites is eliminated
\figref{fig:size_slippery}. In the perfectly slippery limit,
increasing $l_{\rm obst}$ effectively lowers the number of binding
sites: tracers experience the binding energy change only when binding
to obstacle edge sites, but can move freely through obstacle interior
sites. Therefore, $D^*$ and $\alpha_{\rm min}$ increase with obstacle
size, an effect that is larger for higher filling fraction, because
obstacle overlaps at high filling fraction lower the fraction of
obstacles that impede motion and cage tracers. In nearly all cases,
$t_a$ increases with obstacle size, because the effective cage size
grows. The exception occurs for impenetrable obstacles, where
increasing $l_{\rm obst}$ increases the size of vacancies between
cages, allowing caged tracers to escape more quickly.

\section{ Conclusion }

In this paper, we have studied a lattice model of tracer particles
that diffuse and experience crowding due to immobile obstacles. While
most previous work has considered hard (impenetrable) obstacles, we
consider soft (penetrable) obstacles characterized by a binding free
energy that allows tracers to overlap with obstacles. We also consider
the effects of varying the tracer mobility while bound, including the
limiting cases of `sticky' obstacles (which immobilize bound tracers)
and `slippery' obstacles (which allow full tracer mobility), as well as
the intermediate regime between the two.

In some cases, diffusion crowded media leads to dynamics that are
anomalous ($r^2 \sim t^\alpha$) with a constant $\alpha$
\cite{hofling_anomalous_13}.  However, our system typically does not
give a power-law dependence of the MSD on time delay; this has been
seen by others \cite{ellery_modeling_16, jeon_protein_16}.  As a
result, we quantified a long-time diffusion constant ($D^*$), the
timescale on which the systems transitions from anomalous to Fickian
($t_a$), and the minimum instantaneous anomalous exponent
($\alpha_{\rm min}$).

Our results demonstrate the key differences between sticky and
slippery obstacles.  For sticky obstacles, increasing the obstacle
filling fraction decreases the diffusion coefficient and increases the
degree of anomalous diffusion. Above an upper critical occupancy
$ \nu^u \approx 0.72 $ in 2D, diffusion becomes anomalous at all
times, independent of binding energy.  In the sticky case, the minimum
anomalous exponent, $ \alpha_{\rm min} $ monotonically decreases with
filling fraction, because adding more obstacles creates more cages in
which tracers become transiently confined.

For slippery obstacles, by contrast, tracers always reach normal
diffusion after a sufficiently long time; even increasing the filling
fraction above the percolation threshold does not eliminate tracer
motion. For nonzero binding free energy, we find a novel non-monotonic
dependence of $D^*$ on filling fraction: increasing the filling
fraction away from zero introduces binding sites that slow tracer
diffusion, but for sufficiently high filling fraction, bound mobility
allows tracer motion along clusters of obstacles.  The anomalous
exponent decreases with binding energy magnitude, but varies
non-monotonically with filling fraction. For low filling fraction,
$\alpha_{\rm min}$ decreases as more obstacles are added, because
binding transiently traps tracers on isolated obstacles. For
sufficiently high density, diffusion becomes more normal when tracers
hop along clusters of obstacles while bound.

For intermediate `semi-slippery' obstacles, we demonstrate that in the
crossover from from sticky to slippery behavior, $D^*$,
$\alpha_{\rm min}$, and $t_a$ vary smoothly. Increasing bound
diffusion always makes the diffusion coefficient larger and the
diffusive motion less anomalous.

We varied obstacle size to examine how relatively large obstacle
`domains' affects tracer motion in our model. For sticky obstacles,
increasing obstacle size above 1 led to a sharp jump in tracer
properties. This occurs because larger obstacles always contain
interior obstacle sites, which are inaccessible to tracers in the
sticky model.  For large obstacles, increasing repulsive binding
energy tends to increase the tracer diffusion coefficient, because
tracers spend less time trapped in a binding site.
 
For slippery obstacles, perimeter and interior obstacle sites are both
accessible, which means that varying obstacle size has effects that
are easier to understand intuitively. The diffusion coefficient and
anomalous exponent increase with obstacle size, because larger
obstacles lead to a fewer obstacle-empty boundaries.  The effect of
obstacle size on $t_a$ varied with filling fraction, due to competing
effects on increasing cage size and increasing gaps between cages.

Our models separately represent effects of soft interactions (through
the binding energy) and bound-state motion (through obstacle
stickiness/slipperiness).  Sticky and slippery obstacles show
dramatically different tracer dynamics, even at short time and low
filling fraction. Slippery obstacles lead to a diffusion coefficient
which varies non-monotonically with filling fraction, with high values
at both high and low obstacle densities.  As the filling fraction
increases from zero, the particles are more and more inhibited by
obstacles.  However, as the obstacle density increases, particles
which bind can more easily move between obstacles.  This may describe transport
factor motion within the nuclear pore complex, where transport factors
can slide on the disordered FG Nups
\cite{raveh_slideandexchange_16}. Therefore, biological systems may
use soft interactions and slippery obstacles to allow particle
diffusion, even in the highly crowded cellular interior.

Our work highlights how soft interactions\ and bound-state mobility
can dramatically change tracer motion. These effects are relevant to
biological systems, ranging from membrane-less organelles to lipid
rafts.  Although most previous theoretical work on crowded diffusion
has focused on the anomalous exponent, these biological examples
highlight the importance of changes in the diffusion coefficient. For
example, proteins which do not passage through the nuclear pore
complex on biologically relevant time scales (minutes to hours) cannot
have biological effects, and so the speed of passage is the
fundamentally important biological quantity. The long-time diffusion
coefficient varies dramatically in our model between hard obstacles,
soft sticky obstacles and soft slippery obstacles (figure
\ref{fig:comprDT}).  Thus, the effective permeability of obstacles and
the degree to which bound particles can diffuse can be used by cells
to tune macromolecular motion.

\section*{Acknowledgments}

We would like to thank Matthew A. Glaser, Hui-Shun Kuan, and Jeffrey
M. Moore for thoughtful discussions.  This work was funded by NSF
grants DMR-1551095 (MDB) and DMR-1420736 (MWS), and NIH grants
K25GM110486 (MDB) and R35GM119755 (LEH).  The authors acknowledge the
Biofrontiers Computing Core at the University of Colorado Boulder for
providing High Performance Computing resources (NIH 1S10OD012300)
supported by Biofrontiers IT.

\section*{References }

\bibliography{NPC.bib}
\bibliographystyle{unsrt.bst}
\end{document}